\documentclass[11pt]{article}
\usepackage{amsmath,amssymb,amsthm,amsxtra,overpic,bbm,bm,epsfig}
\def\thefootnote{\fnsymbol{footnote}}

\begin{document}

\vspace{0.2cm}

\begin{center}
{\Large\bf Leptonic commutators and clean T violation in neutrino
oscillations}
\end{center}

\vspace{0.1cm}

\begin{center}
{\bf Zhi-zhong Xing}
\footnote{E-mail: xingzz@ihep.ac.cn} \\
{Institute of High Energy Physics, Chinese Academy of
Sciences, P.O. Box 918, Beijing 100049, China \\
Theoretical Physics Center for Science Facilities, Chinese Academy
of Sciences, Beijing 100049, China
}
\end{center}

\vspace{1.5cm}

\begin{abstract}
A realistic medium- or long-baseline neutrino experiment
may suffer from terrestrial matter effects which are likely to
contaminate the genuine T-violating asymmetry between $\nu^{}_\alpha
\to \nu^{}_\beta$ and $\nu^{}_\beta \to \nu^{}_\alpha$ oscillations.
With the help of the commutators of lepton mass matrices in matter,
we show that this kind of contamination is negligible for a variety
of experiments provided the neutrino beam energy $E$ and the
baseline length $L$ satisfy the condition $10^{-7} (L/{\rm km})^2
({\rm GeV}/E) \ll 1$. The same observation is true for the
CP-violating term of the asymmetry between $\nu^{}_\alpha \to \nu^{}_\beta$
and $\overline{\nu}^{}_\alpha \to \overline{\nu}^{}_\beta$ oscillations in
matter.
\end{abstract}

\begin{flushleft}
\hspace{0.8cm} PACS number(s): 14.60.Pq, 13.10.+q, 25.30.Pt \\
\hspace{0.8cm} Keywords: leptonic commutators, neutrino
oscillations, T violation, CP violation, matter effects
\end{flushleft}

\def\thefootnote{\arabic{footnote}}
\setcounter{footnote}{0}

\newpage

In the framework of three neutrino flavors, the strength of leptonic
CP and T violation in neutrino oscillations is measured by a
universal and rephasing-invariant parameter ${\cal J}$ \cite{J}
defined through
\begin{eqnarray}
{\rm Im} \left(V^{}_{\alpha i} V^{}_{\beta j} V^*_{\alpha j} V^*_{\beta i}
\right) = {\cal J} \sum_{\gamma} \epsilon^{}_{\alpha\beta\gamma}
\sum^{}_k \epsilon^{}_{ijk} \; ,
\end{eqnarray}
in which $V$ is the $3\times 3$ unitary lepton flavor mixing matrix
\cite{MNSP}, the Greek subscripts run over $(e, \mu, \tau)$ and the
Latin subscripts run over $(1, 2, 3)$. In terms of the standard
parametrization of $V$ \cite{PDG}, we have ${\cal J} =
\sin\theta^{}_{12} \cos\theta^{}_{12} \sin\theta^{}_{23}
\cos\theta^{}_{23} \sin\theta^{}_{13} \cos^2\theta^{}_{13}
\sin\delta$. The fact that the smallest flavor mixing angle
$\theta^{}_{13}$ is not that small \cite{DYB} makes it very hopeful
to see appreciable CP- and T-violating effects in the lepton sector
provided the unknown Dirac phase $\delta$ is not strongly suppressed
\cite{Xing12}. The Jarlskog parameter ${\cal J}$ is directly
associated with the T-violating asymmetry between the probabilities
of $\nu^{}_\alpha \to \nu^{}_\beta$ and $\nu^{}_\beta \to
\nu^{}_\alpha$ oscillations \cite{Cabibbo}:
\begin{eqnarray}
{\cal A}^{}_{\rm T} = P(\nu^{}_\alpha \to \nu^{}_\beta) -
P(\nu^{}_\beta \to \nu^{}_\alpha) = 16 {\cal J}
\sin\frac{\Delta^{}_{12} L}{4E} \sin\frac{\Delta^{}_{23} L} {4E}
\sin\frac{\Delta^{}_{31} L}{4E} \; ,
\end{eqnarray}
in which $(\alpha, \beta) = (e, \mu)$, $(\mu, \tau)$ or $(\tau, e)$,
$\Delta^{}_{ij} \equiv m^2_i - m^2_j$ with $m^{}_i$ or $m^{}_j$
being the neutrino masses (for $i, j = 1, 2, 3$), $E$ denotes the
neutrino beam energy, and $L$ is the distance between the neutrino
source and the detector. In a realistic medium- or long-baseline
neutrino oscillation experiment, however, the terrestrial matter
effects may contaminate the measurement and thus affect the
determination of ${\cal J}$ (or equivalently, $\delta$). It is
generally expected that the T-violating asymmetry between
$\nu^{}_\alpha \to \nu^{}_\beta$ and $\nu^{}_\beta \to
\nu^{}_\alpha$ oscillations should be less sensitive to matter
effects than the CP-violating asymmetry between $\nu^{}_\alpha \to
\nu^{}_\beta$ and $\overline{\nu}^{}_\alpha \to
\overline{\nu}^{}_\beta$ oscillations \cite{Kuo}, because matter
effects on the CP-conserving parts of $P(\nu^{}_\alpha \to
\nu^{}_\beta)$ and $P(\nu^{}_\beta \to \nu^{}_\alpha)$ are exactly
the same. Therefore, we have the T-violating asymmetry in matter
\cite{Ohlsson}:
\begin{eqnarray}
\tilde{\cal A}^{}_{\rm T} = \tilde{P}(\nu^{}_\alpha \to
\nu^{}_\beta) - \tilde{P}(\nu^{}_\beta \to \nu^{}_\alpha) = 16
\tilde{\cal J} \sin\frac{\tilde{\Delta}^{}_{12} L}{4E}
\sin\frac{\tilde{\Delta}^{}_{23} L} {4E}
\sin\frac{\tilde{\Delta}^{}_{31} L}{4E} \; ,
\end{eqnarray}
where $\tilde{\Delta}^{}_{ij} \equiv \tilde{m}^2_{i} -
\tilde{m}^2_j$ with $\tilde{m}^{}_i$ or $\tilde{m}^{}_j$ being the
effective neutrino masses in matter, and a ``tilde" always indicates
the quantities contaminated by matter. It is known that $\tilde{\cal
J} \tilde{\Delta}^{}_{12} \tilde{\Delta}^{}_{23}
\tilde{\Delta}^{}_{31} = {\cal J} \Delta^{}_{12} \Delta^{}_{23}
\Delta^{}_{31}$ holds for a constant matter profile \cite{Naumov},
but under what condition can $\tilde{\cal A}^{}_{\rm T} = {\cal
A}^{}_{\rm T}$ hold to a good degree of accuracy?

Of course, one may trivially argue that $\tilde{\cal A}^{}_{\rm T}$
must be approximately equal to ${\cal A}^{}_{\rm T}$ if the
terrestrial matter effects are {\it small}. But then one has to specify
{\it how small} is small. In the literature some authors have empirically
answered the above question by doing a careful numerical calculation of
$\tilde{\cal A}^{}_{\rm T}$ for a given matter density profile and examining
its deviation from ${\cal A}^{}_{\rm T}$ \cite{Ohlsson}.

The present paper aims to answer the same question in a less numerical
but more analytical way, with the help of the commutators of lepton mass
matrices developed in Ref. \cite{J2}. We shall specify the conditions
under which $\tilde{\cal A}^{}_{\rm T} = {\cal A}^{}_{\rm T}$ hold
either exactly or to a good degree of accuracy with no need of doing
a lot of numerical calculations. Our work is
different from Ref. \cite{J2} at least in two aspects: (a) we follow
a more concise procedure, in which some complication and ambiguity
can be avoided, to establish the relationship between $\tilde{\cal
A}^{}_{\rm T}$ and ${\cal A}^{}_{\rm T}$; (b) we figure out the
explicit condition to assure $\tilde{\cal A}^{}_{\rm T} \simeq {\cal
A}^{}_{\rm T}$, which can be used to judge a realistic medium- or
long-baseline neutrino oscillation experiment. In comparison with
Ref. \cite{Ohlsson} and other previous works, our present study
provides not only an alternative approach for understanding terrestrial
matter effects on CP and T violation in neutrino oscillations but also
an instructive application of the language of leptonic commutators.

\vspace{0.3cm}

Let us work in the flavor basis where the charged lepton mass matrix
$M^{}_\ell$ is diagonal (i.e., $M^{}_\ell = D^{}_\ell \equiv {\rm
Diag}\{m^{}_e, m^{}_\mu, m^{}_\tau\}$). Then the neutrino mass
matrix $M^{}_\nu$ can be written as $M^{}_\nu = V D^{}_\nu
V^\dagger$ with $D^{}_\nu \equiv {\rm Diag}\{m^{}_1, m^{}_2,
m^{}_3\}$. In matter, we denote the effective neutrino mass matrix
as $\tilde{M}^{}_\nu$ and the effective lepton flavor mixing matrix
as $\tilde{V}$. Then we have $\tilde{M}^{}_\nu = \tilde{V}
\tilde{D}^{}_\nu \tilde{V}^\dagger$ with $\tilde{D}^{}_\nu \equiv
{\rm Diag}\{\tilde{m}^{}_1, \tilde{m}^{}_2, \tilde{m}^{}_3\}$. In a
way similar to Ref. \cite{J2}, we define the leptonic  commutators
in vacuum and in matter:
\begin{eqnarray}
X^{}_\xi \hspace{-0.2cm} & \equiv & \hspace{-0.2cm} \left[M^{}_\ell
M^\dagger_\ell, \exp\left(2 {\rm i} \xi M^{}_\nu
M^\dagger_\nu\right) \right] = \left[D^2_\ell, \exp\left(2 {\rm i}
\xi V D^2_\nu V^\dagger\right) \right] \; ,
\nonumber \\
\tilde{X}^{}_\xi \hspace{-0.2cm} & \equiv & \hspace{-0.2cm}
\left[M^{}_\ell M^\dagger_\ell, \exp\left(2 {\rm i} \xi
\tilde{M}^{}_\nu \tilde{M}^\dagger_\nu \right) \right] =
\left[D^2_\ell, \exp\left( 2 {\rm i} \xi \tilde{V} \tilde{D}^2_\nu
\tilde{V}^\dagger\right) \right] \; ,
\end{eqnarray}
where $\xi = L/(4E)$ with $E$ being the neutrino beam energy and $L$
being the baseline length. When a neutrino beam travels through
a medium, it sees a nontrivial refractive index because of the
coherent forward scattering from the constituents of the medium via
the charged current interactions \cite{M}. The effective Hamiltonian
responsible for the propagation of neutrinos in normal matter can be
expressed as
\begin{eqnarray}
{\cal H}^{}_{\rm m} = \frac{1}{2E} \tilde{V} \tilde{D}^2_\nu
\tilde{V}^\dagger = \frac{1}{2E} \left(V D^2_\nu V^\dagger +
T^{}_{\rm m} \right) \; ,
\end{eqnarray}
where $T^{}_{\rm m} = {\rm Diag}\{A, 0, 0\}$ with $A = 2\sqrt{2}
G^{}_{\rm F} N^{}_e E$. Here $G^{}_{\rm F}$ is the Fermi constant
and $N^{}_e$ denotes the background density of electrons. Eq. (5)
implies a simple relationship between $\tilde{M}^{}_\nu
\tilde{M}^\dagger_\nu$ and $M^{}_\nu M^\dagger_\nu$; that is,
$\tilde{V} \tilde{D}^2_\nu \tilde{V}^\dagger = V D^2_\nu V^\dagger +
T^{}_{\rm m}$. This relationship will allow us to establish a
relation between the determinants of $X^{}_\xi$ and
$\tilde{X}^{}_\xi$ under some conditions, as we shall see later on.

The exponential of a square matrix $Z$ is given by the power series
$\exp (Z) = I + Z + Z^2/2! + Z^3/3! + \cdots$, where $I$ denotes the
identity matrix. Because $M^{}_\nu M^\dagger_\nu$ has three
different eigenvalues, we obtain
\begin{eqnarray}
\exp\left(2 {\rm i} \xi V D^2_\nu V^\dagger\right) =
\sum^\infty_{n=0} \frac{\left(2 {\rm i} \xi V D^2_\nu
V^\dagger\right)^n} {n!} = V \left[\sum^\infty_{n=0} \frac{\left(2
{\rm i} \xi D^2_\nu \right)^n} {n!}\right] V^\dagger = V \exp\left(2
{\rm i} \xi D^2_\nu\right) V^\dagger \; .
\end{eqnarray}
The commutator $X^{}_\xi$ turns out to be
\begin{eqnarray}
X^{}_\xi = \left[D^2_\ell, V \exp\left(2 {\rm i} \xi D^2_\nu\right)
V^\dagger \right] = \sum^\infty_{n=0} \frac{\left(2 {\rm i}
\xi\right)^n} {n!} \left[D^2_\ell, V D^{2n}_\nu V^\dagger\right] \;
.
\end{eqnarray}
It is easy to figure out the explicit expression of $\left[D^2_\ell,
V D^{2n}_\nu V^\dagger\right]$, which must be traceless:
\begin{eqnarray}
\left[D^2_\ell, V D^{2n}_\nu V^\dagger\right] = \left[\begin{matrix}
0 & \Delta^{}_{e\mu} \displaystyle \sum^3_{i=1} m^{2n}_i V^{}_{e i}
V^*_{\mu i} & \Delta^{}_{e\tau} \displaystyle \sum^3_{i=1} m^{2n}_i
V^{}_{e i} V^*_{\tau i} \cr \Delta^{}_{\mu e} \displaystyle
\sum^3_{i=1} m^{2n}_i V^*_{e i} V^{}_{\mu i} & 0 &
\Delta^{}_{\mu\tau} \displaystyle \sum^3_{i=1} m^{2n}_i V^{}_{\mu i}
V^*_{\tau i} \cr \Delta^{}_{\tau e} \displaystyle \sum^3_{i=1}
m^{2n}_i V^*_{e i} V^{}_{\tau i} & \Delta^{}_{\tau\mu} \displaystyle
\sum^3_{i=1} m^{2n}_i V^*_{\mu i} V^{}_{\tau i} & 0 \cr
\end{matrix} \right] \; ,
\end{eqnarray}
in which $\Delta^{}_{\alpha \beta} \equiv m^2_\alpha - m^2_\beta$
(for $\alpha, \beta = e, \mu, \tau$). As a result,
\begin{eqnarray}
X^{}_\xi = \left[\begin{matrix} 0 & \Delta^{}_{e\mu} \displaystyle
\sum^3_{i=1} \exp \left(2 {\rm i} \xi m^{2}_i\right) V^{}_{e i}
V^*_{\mu i} & \Delta^{}_{e\tau} \displaystyle \sum^3_{i=1} \exp
\left(2 {\rm i} \xi m^{2}_i\right) V^{}_{e i} V^*_{\tau i} \cr
\Delta^{}_{\mu e} \displaystyle \sum^3_{i=1} \exp \left(2 {\rm i}
\xi m^{2}_i\right) V^*_{e i} V^{}_{\mu i} & 0 & \Delta^{}_{\mu\tau}
\displaystyle \sum^3_{i=1} \exp \left(2 {\rm i} \xi m^{2}_i\right)
V^{}_{\mu i} V^*_{\tau i} \cr \Delta^{}_{\tau e} \displaystyle
\sum^3_{i=1} \exp \left(2 {\rm i} \xi m^{2}_i\right) V^*_{e i}
V^{}_{\tau i} & \Delta^{}_{\tau\mu} \displaystyle \sum^3_{i=1} \exp
\left(2 {\rm i} \xi m^{2}_i\right) V^*_{\mu i} V^{}_{\tau i} & 0 \cr
\end{matrix} \right] \; .
\end{eqnarray}
A straightforward calculation leads us to the determinant of
$X^{}_\xi$ as follows:
\begin{eqnarray}
\det X^{}_\xi \hspace{-0.2cm} & = & \hspace{-0.2cm} 2{\rm i}
\Delta^{}_{e \mu} \Delta^{}_{\mu \tau} \Delta^{}_{\tau e}
\sum^3_{i=1} \sum^3_{j=1} \sum^3_{k=1} {\rm Im} \left(V^{}_{e i}
V^{}_{\mu j} V^{}_{\tau k} V^*_{e k} V^*_{\mu i} V^*_{\tau j}\right)
\exp\left[2 {\rm i} \xi \left( m^2_i + m^2_j + m^2_k \right)\right]
\nonumber \\
\hspace{-0.2cm} & = & \hspace{-0.2cm} -4 {\cal J} \Delta^{}_{e \mu}
\Delta^{}_{\mu \tau} \Delta^{}_{\tau e} \left[ \sin \left(2 \xi
\Delta^{}_{12}\right) + \sin \left(2 \xi \Delta^{}_{23}\right) +
\sin \left(2 \xi \Delta^{}_{31}\right) \right] \exp\left(2 {\rm i}
\xi \sum^3_{i=1} m^2_i \right) \; ,
\end{eqnarray}
where ${\cal J}$ has been defined in Eq. (1). The calculation of
$\det \tilde{X}^{}_\xi$ is exactly analogous, and the final results
of $\det X^{}_\xi$ and $\det \tilde{X}^{}_\xi$ are
\begin{eqnarray}
\det X^{}_\xi \hspace{-0.2cm} & = & \hspace{-0.2cm} 16 {\cal J}
\Delta^{}_{e \mu} \Delta^{}_{\mu \tau} \Delta^{}_{\tau e} \sin
\frac{\Delta^{}_{12} L}{4E} \sin \frac{\Delta^{}_{23} L}{4E} \sin
\frac{\Delta^{}_{31} L}{4E} \exp\left(2 {\rm i} \xi \sum^3_{i=1}
m^2_i \right) \; ,
\nonumber \\
\det \tilde{X}^{}_\xi \hspace{-0.2cm} & = & \hspace{-0.2cm} 16
\tilde{\cal J} \Delta^{}_{e \mu} \Delta^{}_{\mu \tau}
\Delta^{}_{\tau e} \sin \frac{\tilde{\Delta}^{}_{12} L}{4E} \sin
\frac{\tilde{\Delta}^{}_{23} L}{4E} \sin
\frac{\tilde{\Delta}^{}_{31} L}{4E} \exp\left(2 {\rm i} \xi
\sum^3_{i=1} \tilde{m}^2_i \right) \; .
\end{eqnarray}
Comparing Eq. (11) with Eqs. (2) and (3), we arrive at
\begin{eqnarray}
\det X^{}_\xi \hspace{-0.2cm} & = & \hspace{-0.2cm} {\cal A}^{}_{\rm
T} \Delta^{}_{e \mu} \Delta^{}_{\mu \tau} \Delta^{}_{\tau e}
\exp\left(2 {\rm i} \xi \sum^3_{i=1} m^2_i \right) \; ,
\nonumber \\
\det \tilde{X}^{}_\xi \hspace{-0.2cm} & = & \hspace{-0.2cm}
\tilde{\cal A}^{}_{\rm T} \Delta^{}_{e \mu} \Delta^{}_{\mu \tau}
\Delta^{}_{\tau e} \exp\left(2 {\rm i} \xi \sum^3_{i=1}
\tilde{m}^2_i \right) \; .
\end{eqnarray}
The next step is to establish a relationship between $\det X^{}_\xi$
and $\det \tilde{X}^{}_\xi$, in order to establish a relationship
between ${\cal A}^{}_{\rm T}$ and $\tilde{\cal A}^{}_{\rm T}$. Let
us now come back to Eq. (5).

Given two square matrices $Y$ and $Z$, the exponential of $Y+Z$ can
be expressed as the Zassenhaus expansion \cite{Casas}:
$\exp\left(Y+Z\right) = \exp\left(Y\right) \exp\left(Z\right)
\exp\left(-\left[Y, Z\right]/2\right) \exp\left(\left[Z, \left[Y,
Z\right] \right]/3 + \left[Y, \left[Y, Z\right] \right]/6\right)
\cdots$. We see that $\exp\left(Y+Z\right) = \exp\left(Y\right)
\exp\left(Z\right)$ exactly holds if $Y$ and $Z$ are commutable
(i.e., $YZ = ZY$). In our case, what we are concerned about is the
commutator
\begin{eqnarray}
\frac{\left(2 {\rm i} \xi\right)^2}{2} \left[V D^2_\nu V^\dagger ,
T^{}_{\rm m}\right] = 2 \xi^2 A \left[\begin{matrix} 0 &
\displaystyle \sum^3_{i=2} \Delta^{}_{i1} V^{}_{e i} V^*_{\mu i} &
\displaystyle \sum^3_{i=2} \Delta^{}_{i1} V^{}_{e i} V^*_{\tau i}
\cr - \displaystyle \sum^3_{i=2} \Delta^{}_{i1} V^*_{e i} V^{}_{\mu
i} & 0 & 0 \cr - \displaystyle \sum^3_{i=2} \Delta^{}_{i1} V^*_{e i}
V^{}_{\tau i} & 0 & 0 \cr \end{matrix} \right ] \; .
\end{eqnarray}
If the condition
\begin{eqnarray}
\gamma^{}_{\rm m} \equiv \xi^2 A
\left[\left|\sum^3_{i=2} \Delta^{}_{i1} V^{}_{e i} V^*_{\mu
i}\right| + \left|\sum^3_{i=2} \Delta^{}_{i1} V^{}_{e i} V^*_{\tau
i}\right|\right] \ll 1
\end{eqnarray}
is satisfied, one may argue that the commutator in Eq. (13) is far
smaller than the identity matrix and thus
\begin{eqnarray}
\exp\left( 2 {\rm i} \xi \tilde{V} \tilde{D}^2_\nu
\tilde{V}^\dagger\right) \simeq \exp\left(2 {\rm i} \xi V D^2_\nu
V^\dagger\right) \exp\left(2 {\rm i} \xi T^{}_{\rm m}\right) \;
\end{eqnarray}
holds as a good approximation. Note that $\exp \left(2 {\rm i} \xi
D^2_\nu \right) = {\rm Diag} \left\{\exp\left(2 {\rm i} \xi
m^2_1\right), \exp\left(2 {\rm i} \xi m^2_2\right), \exp\left(2 {\rm
i} \xi m^2_1\right)\right\}$ and $\exp \left(2 {\rm i} \xi T^{}_{\rm
m} \right) = {\rm Diag} \left\{\exp\left(2 {\rm i} \xi A\right), 1,
1\right\}$ hold. Therefore,
\begin{eqnarray}
\exp\left(2 {\rm i} \xi \tilde{V} \tilde{D}^2_\nu \tilde{V}^\dagger
\right) \simeq \left[ \begin{matrix} \displaystyle \sum^3_{i=1}
\exp\left[2 {\rm i} \xi \left(m^2_i + A\right) \right] |V^{}_{e
i}|^2 & \displaystyle \sum^3_{i=1} \exp\left(2 {\rm i} \xi
m^2_i\right) V^{}_{e i} V^*_{\mu i} & \displaystyle \sum^3_{i=1}
\exp\left(2 {\rm i} \xi m^2_i\right) V^{}_{e i} V^*_{\tau i} \cr
\displaystyle \sum^3_{i=1} \exp\left[2 {\rm i} \xi \left(m^2_i +
A\right) \right] V^*_{e i} V^{}_{\mu i} & \displaystyle \sum^3_{i=1}
\exp\left(2 {\rm i} \xi m^2_i\right) |V^{}_{\mu i}|^2 &
\displaystyle \sum^3_{i=1} \exp\left(2 {\rm i} \xi m^2_i\right)
V^{}_{\mu i} V^*_{\tau i} \cr \displaystyle \sum^3_{i=1} \exp\left[2
{\rm i} \xi \left(m^2_i + A\right) \right] V^*_{e i} V^{}_{\tau i} &
\displaystyle \sum^3_{i=1} \exp\left(2 {\rm i} \xi m^2_i\right)
V^*_{\mu i} V^{}_{\tau i} & \displaystyle \sum^3_{i=1} \exp\left(2
{\rm i} \xi m^2_i\right) |V^{}_{\tau i}|^2 \cr
\end{matrix} \right] \;
\end{eqnarray}
can be obtained provided $\gamma^{}_{\rm m} \ll 1$. The
commutator $\tilde{X}^{}_\xi$ is then given by
\begin{eqnarray}
\tilde{X}^{}_\xi \simeq \left[\begin{matrix} 0 & \Delta^{}_{e\mu}
\displaystyle \sum^3_{i=1} \exp \left(2 {\rm i} \xi m^{2}_i\right)
V^{}_{e i} V^*_{\mu i} & \Delta^{}_{e\tau} \displaystyle
\sum^3_{i=1} \exp \left(2 {\rm i} \xi m^{2}_i\right) V^{}_{e i}
V^*_{\tau i} \cr \Delta^{}_{\mu e} \displaystyle \sum^3_{i=1} \exp
\left[2 {\rm i} \xi \left(m^{2}_i +A\right)\right] V^*_{e i}
V^{}_{\mu i} & 0 & \Delta^{}_{\mu\tau} \displaystyle \sum^3_{i=1}
\exp \left(2 {\rm i} \xi m^{2}_i\right) V^{}_{\mu i} V^*_{\tau i}
\cr \Delta^{}_{\tau e} \displaystyle \sum^3_{i=1} \exp \left[2 {\rm
i} \xi \left(m^{2}_i + A\right) \right] V^*_{e i} V^{}_{\tau i} &
\Delta^{}_{\tau\mu} \displaystyle \sum^3_{i=1} \exp \left(2 {\rm i}
\xi m^{2}_i\right) V^*_{\mu i} V^{}_{\tau i} & 0 \cr
\end{matrix} \right] \;
\end{eqnarray}
in the same approximation. This allows us to get
\begin{eqnarray}
\det \tilde{X}^{}_\xi \simeq \exp\left( 2 {\rm i} \xi A\right) \det
X^{}_\xi = {\cal A}^{}_{\rm T} \exp\left[2 {\rm i} \xi
\left(\sum^3_{i=1} m^2_i + A\right) \right] = {\cal A}^{}_{\rm T}
\exp\left(2 {\rm i} \xi \sum^3_{i=1} \tilde{m}^2_i \right) \; ,
\end{eqnarray}
where we have used the sum rule \cite{Xing04}
\begin{eqnarray}
\sum^3_{i=1} \tilde{m}^2_i = {\rm tr}\left(\tilde{V} \tilde{D}^2_\nu
\tilde{V}^\dagger\right) = {\rm tr}\left(V D^2_\nu V^\dagger +
T^{}_{\rm m}\right) = \sum^3_{i=1} m^2_i + A \; ,
\end{eqnarray}
which can directly be observed from Eq. (5). Comparing Eq. (18) with
the second equality in Eq. (12), we immediately obtain $\tilde{\cal
A}^{}_{\rm T} \simeq {\cal A}^{}_{\rm T}$ under the condition
$\gamma^{}_{\rm m} \ll 1$. Some discussions are in order.
\begin{itemize}
\item     $\tilde{\cal A}^{}_{\rm T} = {\cal A}^{}_{\rm T}$ exactly holds
if $M^{}_\nu M^\dagger_\nu = V D^2_\nu V^\dagger$ and $T^{}_{\rm m}$
are exactly commutable. The latter possibly implies: (a) there is no
matter effect, $A =0$; or (b) the neutrino masses are exactly
degenerate, $m^{}_1 = m^{}_2 = m^{}_3$; or (c) there is no lepton
flavor mixing, $V = I$. In either case (b) or case (c), of course,
there would be no neutrino oscillation phenomenon.

\item     The order of $\gamma^{}_{\rm m}$ can be estimated as
follows. A global fit of current neutrino oscillation data
\cite{Fit} yields $|\Delta^{}_{31}| \sim 30 \Delta^{}_{21}$,
$|V^{}_{e2} V^*_{\mu 2}| \sim 3 |V^{}_{e3} V^*_{\mu 3}|$,
$|V^{}_{e2} V^*_{\tau 2}| \sim 3 |V^{}_{e3} V^*_{\tau 3}|$, and
$|V^{}_{\mu 3}| \sim |V^{}_{\tau 3}|$. Therefore,
\begin{eqnarray}
\gamma^{}_{\rm m} \hspace{-0.2cm} & \sim & \hspace{-0.2cm} 2\xi^2
A \left|\Delta^{}_{31}\right| \left|V^{}_{e3} V^*_{\mu
3}\right| \sim 10^{-7} \left(\frac{L}{\rm km}\right)^2
\left(\frac{\rm GeV}{E}\right) \; , ~~~~
\end{eqnarray}
where $|\Delta^{}_{31}| \sim 2.4 \times 10^{-3} ~{\rm eV}^2$,
$|V^{}_{e3}| \sim 0.15$, $|V^{}_{\mu 3}| \sim 0.7$ \cite{Fit}, and
$A\sim 2.28\times 10^{-4} {\rm eV}^2 E/{\rm GeV}$ \cite{Shrock} have
typically been input. Given the T2K experiment, one has $L \simeq
295$ km and $E \simeq 0.6$ GeV, and thus $\gamma^{}_{\rm m} \sim
1\%$. As for the NO$\nu$A experiment, $L \simeq 810$ km and $E
\simeq 2$ GeV, leading us to $\gamma^{}_{\rm m} \sim 3\%$. So
$\gamma^{}_{\rm m} \ll 1$ is satisfied in both of these two
experiments. In this sense one may simply use Eq. (14) to make a
judgement, instead of doing a careful numerical analysis.
\end{itemize}
If the condition $\gamma^{}_{\rm m} \ll 1$ is not satisfied, then
the commutator in Eq. (13) and the higher order expansion terms will
contribute an appreciable correction to Eq. (15). In other words,
$\tilde{\cal A}^{}_{\rm T}$ should contain appreciable matter
effects in this case, making it difficult to extract clean ${\cal
J}$ or $\delta$ in this kind of experiments.

\vspace{0.3cm}

The above calculation is subject to a neutrino beam traveling in
matter, but it can directly be extended to the case of an
antineutrino beam traveling in matter. Thanks to CPT invariance, we
have $P(\overline{\nu}^{}_\alpha \to \overline{\nu}^{}_\beta) =
P(\nu^{}_\beta \to \nu^{}_\alpha)$ and
$P(\overline{\nu}^{}_\beta \to \overline{\nu}^{}_\alpha) =
P(\nu^{}_\alpha \to \nu^{}_\beta)$ in vacuum. The T-violating
asymmetry between $\overline{\nu}^{}_\alpha \to
\overline{\nu}^{}_\beta$ and $\overline{\nu}^{}_\beta \to
\overline{\nu}^{}_\alpha$ oscillations is therefore given by ${\cal
A}^\prime_{\rm T} = -{\cal A}^{}_{\rm T}$, as one may easily
read off from Eq. (2). In matter, the corresponding T-violating
asymmetry $\tilde{\cal A}^\prime_{\rm T}$ can be approximately equal
to ${\cal A}^\prime_{\rm T}$ provided the condition $\gamma^{}_{\rm
m} \ll 1$ is also satisfied
\footnote{As for an antineutrino beam in vacuum or in matter, one
may simply make the replacements $V \to V^*$, ${\cal J} \to -{\cal
J}$ and $A \to -A$ in the relevant calculations.}.

Note that the CP-violating asymmetry ${\cal A}^{}_{\rm CP} =
P(\nu^{}_\alpha \to \nu^{}_\beta) - P(\overline{\nu}^{}_\alpha \to
\overline{\nu}^{}_\beta)$ is exactly equal to the T-violating
asymmetry ${\cal A}^{}_{\rm T}$ in vacuum, as guaranteed by CPT
invariance. In matter, however, the asymmetry $\tilde{\cal A}^{}_{\rm CP} =
\tilde{P}(\nu^{}_\alpha \to \nu^{}_\beta) -
\tilde{P}(\overline{\nu}^{}_\alpha \to \overline{\nu}^{}_\beta)$ is
in general different from $\tilde{\cal A}^{}_{\rm T}$ because the
CP-conserving parts of $\tilde{P}(\nu^{}_\alpha \to \nu^{}_\beta)$
and $\tilde{P}(\overline{\nu}^{}_\alpha \to
\overline{\nu}^{}_\beta)$ are not identical with each other due to
the opposite-sign matter corrections
\footnote{Namely, the effective Hamiltonian responsible for the
propagation of antineutrinos in normal matter is given by ${\cal
H}^\prime_{\rm m} = \left(V^* D^2_\nu V^T - T^{}_{\rm m}
\right)/\left(2 E\right)$, in contrast to ${\cal H}^{}_{\rm m} =
\left(V D^2_\nu V^\dagger + T^{}_{\rm m} \right)/\left(2 E\right)$
for neutrinos.}.
This kind of {\it fake} CPT violation is actually a pure matter
effect \cite{CPT}, which may more or less contaminate the extraction
of the genuine CP-violating effect from $\tilde{\cal A}^{}_{\rm CP}$
\cite{matter}. As for the CP-violating term in the expression of
$\tilde{\cal A}^{}_{\rm CP}$, however, the result obtained
in Eq. (14) or Eq. (20) is also applicable.

Of course, it is always possible to make an analytical expansion of
$\tilde{P}(\nu^{}_\alpha \to \nu^{}_\beta)$ in terms of small
parameters (e.g., the ratio $\Delta^{}_{21}/\Delta^{}_{31}$ and
$\sin\theta^{}_{13}$) \cite{Ohlsson,Sato}, so as to see the sensitivities
of its CP-conserving (or T-conserving) and CP-violating (or
T-violating) parts to terrestrial matter effects. In the present
work we have provided an alternative way towards understanding
matter effects on the T-violating asymmetry in neutrino
oscillations. This way is certainly more compact, although its
usefulness is limited to the cases in which $\gamma^{}_{\rm m} \ll
1$ must be satisfied. Fortunately, it seems that most
currently-proposed medium- and long-baseline neutrino oscillation
experiments are consistent with this condition.

\vspace{0.3cm}

In summary, the discovery of leptonic CP or T violation will be one
of the major targets in tomorrow's experimental neutrino physics. To
achieve this important goal, a number of long- or medium-baseline
neutrino oscillation experiments have been proposed or are under
construction. We have shown that most of such experiments satisfy
the condition $10^{-7} (L/{\rm km})^2 ({\rm GeV}/E) \ll 1$, implying
that the T-violating asymmetry between $\nu^{}_\alpha \to
\nu^{}_\beta$ and $\nu^{}_\beta \to \nu^{}_\alpha$ oscillations
(or between $\overline{\nu}^{}_\alpha \to \overline{\nu}^{}_\beta$
and $\overline{\nu}^{}_\beta \to \overline{\nu}^{}_\alpha$ oscillations)
is essentially free from terrestrial matter effects. In our proof we
have adopted the language of leptonic commutators in matter, which
can provide us with a concise and novel way of understanding
matter effects. We hope that this useful theoretical language could
find some more phenomenological applications in neutrino physics.

\vspace{0.5cm}

I am grateful to Y.F. Li and Y.L. Zhou for stimulating and
useful discussions, and to E.Kh. Akhmedov and T. Ohlsson for
helpful communications. This work is supported in part by the National
Natural Science Foundation of China under Grant No. 11135009.

\end{document}